\documentclass[onecolumn,12pt]{article}
\usepackage{graphicx}
\begin{document}
\title{Neutrino Mass Matrix from Seesaw Mechanism Subjected to Texture Zero and Invariant Under a Cyclic Permutation}
\date{}
\maketitle

\begin{center}
\textbf{Asan Damanik}\footnote{E-mail:d.asan@lycos.com}\\
Department of Physics, Sanata Dharma University, \\ Kampus III USD Paingan, Maguwoharjo, Sleman, Yogyakarta, Indonesia\\
and Department of Physics, Gadjah Mada University,\\  Bulaksumur, Yogyakarta, Indonesia.\\
\end{center}
\begin{center}
\textbf{Mirza Satriawan} and \textbf{Muslim}\\
Department of Physics, Gadjah Mada University,\\  Bulaksumur, Yogyakarta, Indonesia.\\
\end{center}
\begin{center}
\textbf{Pramudita Anggraita}\\
National Nuclear Energy Agency (BATAN), Jakarta, Indonesia.
\end{center}

\abstract{We evaluate the predictive power of the neutrino mass matrices arising from seesaw mechanism subjected to texture zero and satisfying a cyclic permutation invariant.  We found that only  two from eight possible patterns of the neutrino mass matrices to be invariant under a cyclic permutation.  The two resulted neutrino mass matrices which are invariant under a cyclic permutation can be used qualitatively to explain the neutrino mixing phenomena for solar neutrino and to derive the mixing angle that agrees with the experimental data.}

\section{Introduction}
For more than two decades the solar neutrino flux measured on Earth has been much less than predicted by solar model \cite{Pantaleone91}. The solar neutrino deficit can be explained if the neutrino undergoes oscillation during its propagation to earth.  Neutrino oscillation is the change of neutrinos flavor during neutrinos propagation from one place to another. The neutrino oscillation implies that the neutrinos have a non-zero mass or at least one of the three neutrino flavors as we have already known today has non-zero mass and some mixing does exist in neutrino sector.
Recently, there is a convincing evidence that the neutrinos have a non-zero mass.  This evidence was based on the experimental facts that both solar and atmospheric neutrinos undergoing a change from one kind of flavor to another one during the neutrinos propagation in vacuum or matter \cite{Fukuda98,Fukuda99, Ahn03,Toshito01, Giacomelli01, Ahmad02}. These facts are in contrast to the Standard Model of Particle Physics, especially Electro-weak interaction which is based on $SU(2)_{L}\otimes U(1)_{Y}$ gauge, that is neutrinos are massless.

A global analysis of neutrino oscillations data gives the best fit value to solar neutrino mass-squared differences \cite{Gonzales-Garcia04}:
\begin{eqnarray}
\Delta m_{21}^{2}=(8.2_{-0.3}^{+0.3})\times 10^{-5}~eV^2~
\end{eqnarray}
with
\begin{eqnarray}
\tan^{2}\theta_{21}=0.39_{-0.04}^{+0.05},
\end{eqnarray}
and for the atmospheric neutrino mass-squared differences
\begin{eqnarray}
\Delta m_{32}^{2}=(2.2_{-0.4}^{+0.6})\times 10^{-3}~eV^2~
\end{eqnarray}
with
\begin{eqnarray}
\tan^{2}\theta_{32}=1.0_{-0.26}^{+0.35},
\end{eqnarray}
where $\Delta m_{ij}^2=m_{i}^2-m_{j}^2~ (i,j=1,2,3)$ with $m_{i}$ as the neutrino mass eigenstates basis $\nu_{i}~(i=1,2,3)$ and $\theta_{ij}$ is the mixing angle between $\nu_{i}$ and $\nu_{j}$.  The mass eigenstates related to weak (flavor) eigenstates basis $(\nu_{e},\nu_{\mu},\nu_{\tau})$ is as follows
\begin{eqnarray}
\bordermatrix{& \cr
&\nu_{e}\cr
&\nu_{\mu}\cr
&\nu_{\tau}\cr}=V\bordermatrix{& \cr
&\nu_{1}\cr
&\nu_{2}\cr
&\nu_{3}\cr}
\end{eqnarray}
where $V$ is the mixing matrix.

To accommodate a non-zero neutrino mass-squared differences and the neutrino mixing, several models for neutrino mass together with the neutrino mass generation have been proposed \cite{Mohapatra98,Akhmedov99,He03,Zee03,Fukugita03,Altarelli04,Dermisek04,Ma03}.  One of the interesting mechanism to generate neutrino mass is the seesaw mechanism, in which the right-handed neutrino $\nu_{R}$ has a large Majorana mass $M_{N}$ and the left-handed neutrino $\nu_{L}$ is given a mass through leakage of the order of $~(m/M)$ with $m$ the Dirac mass \cite{Fukugita03}.  Seesaw mechanism explains not only the smallness of neutrino mass in the electro-weak energy scale but also could account for the large mixing angle in neutrino sector \cite{Gell-Mann79,Tsujomoto05}.  The mass matrix model of a massive Majorana neutrino $M_{N}$ which is constrained by the solar and atmospheric neutrinos deficit and incorporate the seesaw mechanism and Peccei-Quinn symmetry have been reported by Fukuyama and Nishiura \cite{Fukuyama97}.

In this paper, we construct the neutrino mass matrices arise from seesaw mechanism subjected to texture zero and invariant under a cyclyc permutation.  This paper is organized as follows: In Section 2, we determine the possible patterns for the heavy neutrino mass matrices $M_{N}$ subjected to texture zero and then check its invariance under a cyclic permutation.  The resulted $M_{N}$ matrices to be used to obtain the neutrino mass matrices $M_{\nu}^{'}$ arising from seesaw mechanism.  In Section 3, we discuss the predictive power of the resulted neutrino mass matrices $M_{\nu}^{'}$ against the experimental results.  Finally, in Section 4 we give a conclusion.

\section{Texture Zero and Invariant Under a Cyclic Permutation}\label{secII}

According to the seesaw mechanism \cite{Gell-Mann79}, the neutrino mass matrix $M_{\nu}$ is given by
\begin{eqnarray}
M_{\nu}\approx -M_{D}M_{N}^{-1}M_{D}^T
\end{eqnarray}
where $M_{D}$ and $M_{N}$ are the Dirac and Majorana mass matrices respectively.  If we take $M_{D}$ to be diagonal, then the pattern of the neutrino mass matrix $M_{\nu}$ depends only on the pattern of the $M_{N}$ matrix.  From Eq. (5), one can see that the pattern of the $M_{N}^{-1}$ matrix will be preserved in $M_{\nu}$ matrix when $M_{D}$ matrix is diagonal.

If $M_{N}$ matrix has one or more of its elements to be zero (texture zero), then this implies that $M_{N}^{-1}$ matrix has one or more $2\times 2$ sub-matrices with zero determinants \cite{Leontaris96}.  The texture zero of the mass matrix indicates the existence of additional symmetries beyond the Standard Model of Particle Physics.  There are eight possible patterns for $M_{\nu}$ matrices when $M_{N}$ matrix has a texture zero obtained from a seesaw mechanism \cite{Damanik05}.  

Koide \cite{Koide97} have used a vector-like fermions $F_{i}$ in addition to the three families of fermions (leptons and quarks) $f_{i}$ in an $SU(2)_{L}\otimes SU(2)_{R}\otimes U(1)_{Y}$ gauge in order to build a unified mass matrix model for leptons and quarks.  If these fermions and Higgs scalar to be $f_{L}=(2,1), f_{R}=(1,2), F_{L}=(1,1), F_{R}=(1,1), \phi_{L}=(2,1), \phi_{R}=(1,2)$ of $SU(2)_{L}\otimes SU(2)_{R}$ gauge, it implies that the heavy fermions matrix $M_{F}$ has the form
\begin{eqnarray}
M_{F}= {\lambda} m_{0}({\bf1}+3b_{f}X),
\end{eqnarray}
where $b_{f}$ is an $f$-dependent complex parameter, $X$ is a rank-one matrix, $\lambda$ is a constant, $\bf1$ is the identity matrix, and $m_{0}$ satisfy the relation: $m_{L}=m_{R}/\kappa =m_{0}Z$, with $\kappa$ is a constant, and $Z$ is a universal matrix for fermions $f$.
In another Koide's paper \cite{Koide00}, which is related to the neutrino mass matrix following the scheme of seesaw mechanism, he used the form of heavy fermions matrix in Eq. (7) with additional assumption.  The additional assumption is that the form of the mass matrix is invariant under a cyclic permutation   among the fermions $f$.  The form of heavy fermions mass matrix in Eq. (7) can be modified into
\begin{eqnarray}
M_{F}=aE+bS(\theta)
\end{eqnarray}
where $E$ and $S(\theta)$ matrices are given by \cite{Koide00}:
\begin{eqnarray}
E=1/\sqrt{3}\bordermatrix{& & &\cr
&1 &0 &0\cr
&0 &1 &0\cr
&0 &0 &1\cr}
\end{eqnarray}
and
\begin{eqnarray}
S(\theta)=1/\sqrt{6}\bordermatrix{& & &\cr
&0 &e^{i\theta} &e^{-i\theta}\cr
&e^{-i\theta} &0 &e^{i\theta}\cr
&e^{i\theta} &e^{-i\theta} &0\cr}
\end{eqnarray}

To find out the heavy Majorana neutrino mass matrix $M_{N}$, we only need to replace $M_{F}$ by $M_{N}$.  The heavy Majorana neutrinos masses in mass eigenstates basis are the eigenvalues of Eq. (8), and it can be written as
\begin{eqnarray}
m_{1}=1/\sqrt{3}~a+2/\sqrt{6}~b~\cos{\theta}\nonumber\\
m_{2}=1/\sqrt{3}~a-1/\sqrt{6}~b~\cos{\theta}+1/\sqrt{2}~b~
\sin{\theta}
\nonumber\\
m_{3}=1/\sqrt{3}~a-1/\sqrt{6}~b{\cos}{\theta}-1/\sqrt{2}~b
\sin{\theta}
\end{eqnarray}

By taking the $V_{T}$ matrix as 
\begin{eqnarray}
V_{T}=1/\sqrt{3}\bordermatrix{& & &\cr
&1 &1 &1\cr
&\omega &\omega^2 &1\cr
&\omega^2 &\omega &1\cr}
\end{eqnarray}
where $\omega=e^{i2\pi/3}$, the neutrino mass matrix in Eq. (6) could the be written as
\begin{eqnarray}
M_{\nu}^{'}=V_{T}M_{\nu}V_{T}^{T}=-D_{D}(V_{T}^{*}M_{N}V_{T}^{\dagger})^{-1}D_{D}
\end{eqnarray}
where $D_{D}=V_{T}M_{D}V_{T}^{+}=diag(m_{1}^{D},m_{2}^{D},m_{3}^{D})$.  By taking $M_{N}=m_{N}{\bf1}$, and using the relation
\begin{eqnarray}
V_{T}V_{T}^{T}=\bordermatrix{& & &\cr
&1 &0 &0\cr
&0 &0 &1\cr
&0 &1 &0\cr}
\end{eqnarray}
Koide obtained a neutrino mass matrix in flavor basis that can be used to explain the maximal mixing between $\nu_{\mu}$ and $\nu_{\tau}$ which is suggested by the atmospheric neutrino data \cite{Koide00}.

Following Koide's idea accounted in Eq.(8), but taking the form of $V_{T}$to be
\begin{eqnarray}
V_{T}=\bordermatrix{& & &\cr
&-2/\sqrt{6} &1/\sqrt{3} &0\cr
&1/\sqrt{6} &1/\sqrt{3} &1/\sqrt{2}\cr
&1/\sqrt{6} &1/\sqrt{3} &-1/\sqrt{2}\cr}
\end{eqnarray}
such that the $V_{T}$ matrix represents the current experimental data, and assigning texture zero to $M_{N}$ matrix so that it relates to the underlying family symmetry beyond the Standard Model. By using the $V_{T}$ in Eq. (15) we obtained eight possible patterns of the neutrino mass matrices $M_{N}$ with texture zero as one could read in Ref.\cite{Damanik05}. The eight possible $M_{N}$ patterns are:
\begin{eqnarray}
M_{N}=\bordermatrix{& & &\cr
&0 &a &a\cr
&a &b &c\cr
&a &c &b\cr},M_{N}=\bordermatrix{& & &\cr
&a &b &b\cr
&b &c &0\cr
&b &0 &c\cr},M_{N}=\bordermatrix{& & &\cr
&a &b &b\cr
&b &0 &c\cr
&b &c &0\cr},\nonumber \\M_{N}=\bordermatrix{& & &\cr
&a &0 &0\cr
&0 &b &c\cr
&0 &c &b\cr},M_{N}=\bordermatrix{& & &\cr
&0 &a &a\cr
&a &b &0\cr
&a &0 &b\cr},M_{N}=\bordermatrix{& & &\cr
&0 &a &a\cr
&a &0 &b\cr
&a &b &0\cr},\nonumber\\M_{N}=\bordermatrix{& & &\cr
&a &0 &0\cr
&0 &b &0\cr
&0 &0 &b\cr},M_{N}=\bordermatrix{& & &\cr
&a &0 &0\cr
&0 &0 &b\cr
&0 &b &0\cr}.
\end{eqnarray}

By checking the invariant form of the resulting neutrino mass matrices $M_{N}$ with texture zero under a cyclic permutation, we found that there is no $M_{N}$ with texture zero to be invariant under a cyclic permutation.  With additional assumption, there is a possibility to put the $M_{N}$ matrices with texture zero to be invariant under a cyclic permutation, especially for the $M_{N}$ matrices with the patterns:
\begin{eqnarray}
M_{N}=\bordermatrix{& & &\cr
&0 &a &a\cr
&a &0 &b\cr
&a &b &0\cr},
\end{eqnarray}
and
\begin{eqnarray}
M_{N}=\bordermatrix{& & &\cr
&a &0 &0\cr
&0 &b &0\cr
&0 &0 &b\cr}.
\end{eqnarray}

By imposing an additional assumption, that is $a=b$ for both $M_{N}$ matrices in Eqs.(17) and (18), then we obtain two $M_{N}$ matrices to be invariant under a cyclic permutation. The two patterns of the $M_{N}$ matrices which is invariant under a cyclic permutation can be written as follows:
\begin{eqnarray}
M_{N}=m_{N}\bordermatrix{& & &\cr
&0 &1 &1\cr
&1 &0 &1\cr
&1 &1 &0\cr},
\end{eqnarray}
and
\begin{eqnarray}
M_{N}=m_{N}\bordermatrix{& & &\cr
&1 &0 &0\cr
&0 &1 &0\cr
&0 &0 &1\cr}
\end{eqnarray}
where $m_{N}=1/a$. 

By substituting Eqs. (19) and (20) into Eq. (13), we obtain the neutrino mass matrices in flavor basis ($M_{\nu}^{'}$) to be
\begin{eqnarray}
M_{\nu}^{'}=\frac{1}{m_{N}}\bordermatrix{& & &\cr
&(m_{1}^{D})^{2} &0 &0\cr
&0 &(m_{2}^{D})^{2} &0\cr
&0 &0 &(m_{3}^{D})^{2}\cr}
\end{eqnarray}
and
\begin{eqnarray}
M_{\nu}^{'}=\frac {1}{m_{N}}\bordermatrix{& & &\cr
&(m_{1}^{D})^{2} &m_{1}^{D}m_{2}^{D} &m_{1}^{D}m_{3}^{D}\cr
&m_{1}^{D}m_{2}^{D} &(m_{2}^{D})^{2} &m_{2}^{D}m_{3}^{D}\cr
&m_{1}^{D}m_{3}^{D} &m_{3}^{D}m_{2}^{D} &(m_{3}^{D})^{2}\cr}
\end{eqnarray}
respectively. Mohapatra and Rodejohann \cite{Mohapatra06} have also obtained the neutrino mass matrix in Eq.(22) by using the concept of \textit{scaling}.

\section{Discussions}\label{secIII}

Without imposing an additional assumption to the resulting $M_{\nu}$ matrices arising from a seesaw mechanism with texture zero, we have no $M_{\nu}$ matrix to be invariant in form under a cyclic permutation.  
By imposing an additional assumption: $a=b$, we have two $M_{\nu}$ matrices to be invariant under a cyclic permutation.  

By inspecting Eq.(22), one can see that neutrino mass matrix $M_{\nu}^{'}$ arising from the seesaw mechanism subjected to texture zero and invariant in form under a cyclic permutation, could be used to explain the neutrino mixing for both solar and atmospheric neutrinos data.  To extract the predictive power of the resulting neutrino mass matrix in Eq. (22) for the mixing angle $\theta_{21}$, for simplicity, if we pick up the approximation: $(m_{1}^{D})^{2}\approx m_{1}^{D}m_{2}^{D}$, then we can write: $m_{1}^{D}\approx m_{2}^{D}$ and it implies that: $m_{\nu_{e}}\approx m_{\nu_{\mu}}$.  Substituting $m_{1}^{D}\approx m_{2}^{D}$ into Eq. (11), finally we obtain the angle between mass eigenstates $\nu_{1}$ and $\nu_{2}$ to be:
\begin{eqnarray}
\tan(2\theta_{21})=\sqrt{3}
\end{eqnarray}
which corresponds to $\theta_{21}=30^{o}$.  The value of the mixing angle between $\nu_{1}$ and $\nu_{2}$ (certainly between $\nu_{e}$ and $\nu_{\mu}$) is in a good agreement with experimental value as cited in Eq.(2).

The $M_{N}$ matrices with texture zero in the scheme of seesaw mechanism give naturally the neutrino mixing without additional requirement that $M_{N}$ is invariant under a cyclic permutation as proposed by Koide.  This fact can be read in Ref.\cite{Damanik05} for the cases when one, two, and three of the elements of $M_{N}$ matrices to be zero leading to the tri-maximal mixing.  If $M_{N}$ matrix has six of its element to be zero (all of the $M_{N}$ off-diagonal to be zero), then we obtain the same   matrix pattern to $M_{N}$ matrix taken by Koide in his paper. Even though the pattern is similar to that of Koide, the resulting $M_{\nu}^{'}$ matrix is different to Koide's result.  This differences due to the different form of $V_{T}$ between Koide's paper and ours.

\section{Conclusion}\label{secIV}
The eight possible patterns of the $M_{N}$ matrices with texture zero in the seesaw mechanism scheme as can be read in Ref.\cite{Damanik05} could account for the bi- and tri-maximal mixing in neutrino sector without additional requirement that these matrices are invariant in form under a cyclic permutation.   When we impose the requirement to the $M_{N}$ matrices to be invariant in form under a cyclic permutation following Koide's idea, we found that there is no $M_{N}$ matrix to be invariant in form.  But, by imposing an additional assumption, we obtain two of the $M_{N}$ matrices to be invariant in form under a cyclic permutation. One of the two $M_{N}$ matrices which is invariant in form under a cyclic permutation could produces the neutrino mixing mass matrix in flavor basis $M_{\nu}^{'}$.

\section*{Acknowledgments}
The first author would like to thank to the Graduate School Gadjah Mada University Yogyakarta where he is currently a graduate doctoral student, the Dikti Depdiknas for a BPPS Scholarship Program, and the Sanata Dharma University Yogyakarta for granting the study leave and opportunity.

\end{document}